# Unique electronic structure and the shape-change effect of stage-I graphite intercalation compounds with Thorium, Uranium and Plutonium

Kun Yan[1, a)]

[1]State Key Laboratory of Optoelectronic Materials and Technologies, Guangdong Province Key Laboratory of Display Material and Technology, School of Physics, Sun Yat-sen University, Guangzhou 510275, People's Republic of China.

a)Correspondence: yank6@mail2.sysu.edu.cn

## Abstract

GICs doped with elements containing d and f orbitals have been studied rarely. We control the distribution and density of intercalated actinide metals (Th, U and Pu), and consider the effect of changing the distance of two adjacent carbon layers on the electronic structure so as to infer the physical properties of such materials under high pressure or high temperature - which is of great significance in fundamental research. According to band schemas, those GICs are all metallic. The projected density of states (PDOS) indicates that the metal atoms first undergo hybridization of its s, p, d, and f orbitals, and then bond with the carbon's $p_z$ orbitals of the nearest-neighbor octagons. The electron orbital spin up of C and Th is symmetrical with spin down, so there is no electron spin polarization. However, the s, p, d, and f orbitals of U and Pu all exhibit electron spin polarization, which leads to the magnetic properties of the material and make the p orbits of C appear spin polarization around Fermi level. In addition, these three selected elements are the most commonly used raw materials in nuclear fission, so such GICs are expected to become novel nuclear energy storage materials.

**KEYWORDS**

# 1 | Introduction

More than 20 organometallic chemists have been awarded Nobel Prizes since 1963. In the commercial and industrial field, the current market price of high-purity (five 9 to seven 9) organometallic compounds is as high as 10,000 yen per gram, which is an eye-catching and irreplaceable material in electronic industry; its preparation technology is only monopolized in a few developed countries, such as North America, Europe, Japan, etc. As one of the emerging but important organometallic nanomaterials, graphite intercalation compounds (GICs) have begun to penetrate and develop applications in energy, environment, biomedical science, e.g. high-efficiency battery electrodes, supercapacitors, macromolecular detectors, nanosensors, etc. It is worth noting that the rise of the application is precisely benefit from the achievements of fundamental research in the past few decades. To be sure, although GICs have not yet been widely rolled out in industry, this trend is just around the corner.

In previous studies on GICs, superconductivity is one of the focuses, which is based on the special electronic behavior in GICs. In order to increase $T_c$, it is natural to think of intercalating alkali metal or alkaline earth metal atoms between carbon layers, because these elements have s-orbital valence electrons to release, which can increase the charge density of the conjugated electron cloud between the layers. However, so far, the highest $T_c$ of such GICs is still below 20 K - which is a bit unexpected! Compared to the discovered high-temperature superconductor: cuprates ($HgBa_2Ca_2Cu_3O_{8+\delta}$, $T_c$=164 K (under 150 kbar)[1], iron pnictides ($Sm[O_{1-x}F_x]FeAs$, $T_c$=55 K)[2] and the photochemically synthesized ternary carbonaceous sulfur hydride system ($T_c$=288 K, 267 GPa)[3] in 2020, GICs has no obvious competitive advantage. However, plus the correction of the electron-phonon coupling (EPC), the BCS model will give good calculations consistent with experimental results. Increasing the electron concentration of the conjugated cloud or reducing the carbon interlayer space makes the orbital overlap integral larger, produces a larger Fermi surface and new interlayer bands spanning the Fermi level, which is a known means to increase $T_c$ in GICs with alkali or alkaline earth metal elements.

However, the superconductivity of GICs with transition, lanthanide (Ln) and actinide (An) elements containing d- or f-orbital electrons is lacking at present. One reason is the complexity of their electronic structure, which discourages researchers, especially before 2010. Limited by the power of computer, studies based on the first principles seemed to be too difficult.

In this work, I tried to fill that gap. Three An metals were selected with representative valence electron arrangement: Thorium (Th) ($[Rn]6d^27s^2$), Uranium (U) ($[Rn]5f^36d^17s^2$) and Plutonium (Pu) ($[Rn]5f^67s^2$). I controlled the distribution and density of the intercalated metals, and consider the effect of changing carbon interlayers distance $d$ on the electronic structure, so as to infer their feasible properties under extreme conditions (e.g. high pressure). According to band schemas, those GICs are all metallic. The projected density of states (PDOS) indicates that the metal atoms first undergo hybridization of its s, p, d, and f orbitals, and then bond with the carbon's $p_z$ orbitals of the nearest-neighbor octagons. The electron orbital spin up of C and Th is symmetrical with spin down, so there is no electron spin polarization. However, the s, p, d, and f orbitals of U and Pu all exhibit electron spin polarization, which leads to the magnetic properties of the material and make the p orbits of C appear spin polarization around Fermi level. Also, it is verified that the room/distance between the actinide metal atoms and the carbon layers strongly affects the electronic structure of the GICs, which originates from the anisotropy of their p, d, and f orbitals.

## 2 | Model and Methods

Six types of stage-I GICs dipping actinide (An) group metal elements Thorium (Th), Uranium (U) and Plutonium (Pu) were designed. The metal atoms are so large that 2-dimentianal carbon layer is basically tiled with 4+8 polygons[4] (see Fig. 1), and metal atoms bond with carbon in the upper and lower octagons. After geometric optimization, the length of the C-C bond shared by two adjacent octagons is 1.44 Å; the length of the C-C bond shared by a quadrate and an octagon is

1.48 Å. The distance between two adjacent carbon layers is related to the chemical species and distribution feature of the metals (Table 1). When the metal atoms fill the middle of all octagons, its density marks "100%" (see Fig. 1(b)), such as 2DU100, and the distance between adjacent U atoms is 3.45 Å. When the density of metal atoms is "50%" (as shown in Fig. 1(c)), the distance between adjacent U atoms is 4.87 Å. The above data are approximate for the other GICs with Th and Pu, due to the rigidity of the carbon network. In addition, to investigate the "shape-change effect" (changing the distance $d$ between two adjacent carbon layers), I also studied the band and projected density of states (PDOS) when the lattice basis vector $c$ along the Z axis becomes original 90% (noted as "-10%"), 95% ("-5%"), 97.5% ("-2.5%"), 102.5% ("+2.5%"), 105% ("+5%") and 110% ("+10%").

Geometric Optimization and electronic structure were carried out with density functional theory (DFT) as implemented in Vienna Ab initio simulation package (VASP) code. The exchange-correlation potential was treated at the level of generalized gradient approximation (GGA) using Perdew–Burke–Ernzerhof (PBE) functionals. A plane-wave energy cutoff of 520 eV was employed in all calculations. The force and electronic convergence tolerance were set to $10^{-7}$ eV/Å and $10^{-8}$ eV, respectively in the full geometrical optimizations. The k-point grid was set to 10×10×4 in original structure optimization, and 7×7×8 when changing the $c$ (shape-change effect). Also, the k-point grid was set to 20×20×8 in density of state (DOS) calculation and set to 20 in band calculation (More details are shown in Fig. 1(d), (e) and (f)). Since the pure functional method couldn't give a correct band structure corresponding to the PDOS in 2DTh50, we tried hybrid functional method and attained an appropriate result. Spin polarization was performed and due to the locality of d and f electrons a GGA+U approach was used in all calculations. Data was obtained and processed with the help of *vaspkit*[5] and *MATLAB*.

From another point of view, in order to analyze the bonding nature and the electronic behavior between the An metal and the carbon networks, we also calculated the Total Charge Density (TCD), the Charge Density Difference (CDD) [6] and the Electron Localization Function (ELF) [7-12]

## 3 | Results and discussion

### 3.1 | Electronic structures of various metal elements with different densities (50%,

**100%)**

The bands of the six GICs all cross the Fermi surface (Fig.2), so they appear metallic. For Th-doped GICs, the spin-up and spin-down band curves coincide, so there is no spin polarization. The GICs doped with U and Pu show spin polarization. Because the metal with 100% density provides more valence electrons and hybrid orbitals, the band structure is more complex than 50%. 2DU100 has the most complex bands. Compared with Th-doped GICs, U- and Pu-doped GICs have dense bands between (0, -2 eV), which are speculated to be the contribution of U's d orbital or Pu's f orbital through analysis of corresponding PDOS; those dense bands may be so-called "newly formed interlayer bands."

The PDOS of each element shows that the electronic density of states around the Fermi surface is mainly contributed by the $p_z$ orbital electrons of C and the d orbital or f orbital of the metal (M) (see Table 2 for details). In GICs containing U and Pu, the p orbital of carbon has obvious spin polarization at the Fermi surface and in positive energy interval, indicating that the d and f orbits of U or Pu hybridize with the p orbital of C (including $p_x$, $p_y$) to form new orbitals, which are anisotropic. The s and p orbitals of U and Pu also exhibit spin polarization. In addition, the energy intervals in which the d- and f- orbital PDOS distribute are separated. I think the reason is that the extension of the d and f orbitals in space is related to the morphology of the conjugated orbitals and the geometric configuration of GICs (considering EPC). Therefore, some orbitals have high energies under Coulomb action; sharp peaks at various energy positions indicate that these orbitals are discrete.

When the density of metal atoms increases, the discrete energy levels will turn into bands. When the amount of charge transferred from the metal lattice to the carbon network is large enough, the geometric configuration of the carbon network will be locally distorted, and even the nature of the σ bonds (such as bond length) is to change. After structural optimization, the carbon network appear "wrinkled" in all six GICs we discuss (in 2DTh50, the fold reaches 0.43 Å in the z-axis direction), which is similar to that observed experimentally in finite graphene. The reason may be that the carbon network must undergo structural relaxation and release stress, for An elements are too big.

In pristine graphite, the Fermi surface is almost a point where electron-phonon coupling

(EPC) is strongly suppressed. There are three important control parameters for the high conductivity in GICs with chemical formula $MC_x$, x is an experimental value. The first is Z (the valence of M), the second is the atomic mass of the intercalated ion, and the third is the effective mass of the flowing electrons in the interlayer state (m*). Electrons are transferred from intercalated atoms to graphitic bands with two-dimensional feature, or to new three-dimensional (3D) anisotropy bands formed by the hybridization of M's d or f orbitals and graphite's interlayer states. In that model, the fractional factor *f* between 2D or 3D electrons is defined as *Zf:Z(1-f)*. It is confirmed by experimental evidence that, in the reported GICs with superconductivity, the superconducting behavior is mainly contributed by 3D electrons. The three-dimensional electrons experience the mixed electric field of polarized waves generated by the oscillations of positively charged intercalation ions $M^{Z+}$ and negatively charged carbon ions $C^{-fZ/x}$. Both acoustic phonons vibrating perpendicular to the carbon layer and optical phonons vibrating in-plane provide strong phonon-mediated attractive interactions. Electrons on the 2D Fermi surface shield the polar coupling, resulting in a decrease of $T_C$ and an increase of *f*, while intercalated atoms with large *Z* lead to a decrease of *f*. That successfully explains why the $T_C$ of $BaC_6$ (65 mK, 2D-like Fermi surface) is greatly reduced compared to that of $CaC_6$ (11.5 K, 3D-like Fermi surface). T. Sato et al.[13] (2009) found that electrons on a circular Fermi surface at the Γ point are strongly coupled to out-of-plane vibrations. This means that the 3D interlayer orbitals are coupled to the low-frequency phonons of metal and carbon atoms. In my work, such EPCs may be suppressed due to the size and mass of actinides being much larger than carbon. However, the phonons on the triangular Fermi surface at the K point are mainly coupled with in-plane phonons from the σ bonds in the carbon network, so they are high-frequency optical phonons. As predicted by first-principles calculations, the strong interaction of interlayer electrons with the surrounding lattice differs from the picture of a free electron gas. T. Valla et al.[14] (2011) gave the same conclusion that the electrons of the metal atoms are partially transferred to occupy the Fermi surface (FS) (so the larger the area of the FS, the higher the conductivity), and the remaining (electrons) occupy the so-called interlayer states (π* bands), the latter attributed to coupling with "high-frequency phonons in the plane of graphene".

**3.2 | Shape-change effect**

Experimentally, the smaller the distance $d$ between two adjacent carbon layers, the higher the $T_C$. This is because the overlap integral of the hybrid orbitals increases, resulting in more new interlayer bands and more tunnels for electrons to behaviors nearly freely. Both theory and experiment show that electrons are delocalized in the direction parallel to the carbon layer between layers, but localized in the direction perpendicular to the carbon layer; therefore, the near-free electron gas diffuses in the space between the carbon atoms and the metal atoms. Superconductivity is more likely to occur in parallel directions.

Research of the shape-change effect on the electronic structure of GICs is still lack. In some experiments, high pressure was applied to observe changes in physical properties such as $T_C$ and magnetoresistance. Robert P. Smith et al.[15] (2006) found that the superconducting transition temperature $Tc$ increases linearly with pressure for both $C_6Yb$ and $C_6Ca$ with $Tc/dP$ of +0.37±0.01 and +0.50±0.05 K/GPa, respectively. The $T_C$ of $C_6Yb$ first increases with increasing pressure, reaches a peak around 1.8 GPa, and then begins to decrease again. They suggest that the pressure dependence may be due to the plasmon pairing mechanism. This is a typical $T_C$~$P$ relationship. It is generally believed that after the $T_C$ reaches maximum, continuing to pressurize will lead to a phase transition of the lattice, so that the $T_C$ will be different. Further, I think it is the Coulomb repulsion interaction between the electronic states that causes the lattice to deform or even phase transition. Gábor Csanyi et al. [16] (2007) pointed out that the Fermi level increases after doping, accompanied by a decrease and occupation of the dispersive interlayer band.

Based on the same approach, I calculated the electronic structures of the six GICs under the shape-change effect (Fig. 3 to 8). The results are summarized as follows:

1. The qualitative analysis of the valence bond theory is verified from the first-principles calculations.

2. With the change of the basis vector *c* along z-axis, the bands near the Fermi surface change significantly. Compression thins out the bands, while stretching does the opposite (e.g. spin-up band around -1 eV (blue) in 2DU50 and 2DPu100). But GICs remain metallic within the deformation range we consider.

3. The coincidence feature of the PDOS curve conforms to the symmetry requirement of the group to which the GIC's structure belongs. The deformation did not change the symmetry of the group, so the coincidence of the corresponding PDOS curves did not change either.

4. The PDOS curves of carbon's $p_x$ and $p_y$ orbit are coincident and $p_z$ makes a dominant contribution to the Fermi surface. These two phenomena do not change with the crystal deformation. However, $p_x/p_y$ and $p_z$ varies with different metals and deformations.

5. The PDOS curves of the p, d, and f orbitals of the actinide group metals (peak position, intensity, and full width at half maximum, etc.) vary significantly with $d$.

6. The electronic structure of GICs has continuity with the change of $d$, and it is necessary to establish a quantum model to obtain the relationship.

7. With deformation some PDOS becomes smaller, and without deformation it obtains a maximum (as the carbon's p orbit in 2DTh100 shows). Alternatively, the change of some PDOS under tension is continuous, and the same is true under compression; but both are not alike to each other, such as the d orbitals of 2DTh100 and 2DU100. The shape-change effect of 2DPu50 is more pronounced when $c$ is stretched than compressed. The energy distribution of the d orbitals' PDOS in 2DU100 and 2DPu100 will be more concentrated with the stretching of $c$, that is, the corresponding bands will be more dense.

To summarize: The shape-change effect significantly affects the electronic structure near the Fermi surface. When shrinking $c$, the energy band density increases and the bands below the Fermi surface are raised, resulting in more new interlayer bands at the Fermi surface and more bands overlap in the topography; carbon's $p_z$ and the d and f orbitals of the actinide metals contribute to the peak and the total integrated area of the PDOS at the Fermi level. Therefore, the shape-change effect decisively affects the high conductivity parallel to the carbon layer and the kinetic stability of the whole nanomaterial system.

## 3.3 | Total Charge Density (TCD), Charge Density Difference (CDD) and Electron Localization Function (ELF)

The total charge density (TCD), charge density difference (CDD) and electron local function (ELF), can help to vividly and intuitively judge the bonding nature between intercalated atoms and carbon networks. This approach has been widely used in the analysis of solid-state high-density hydrogen storage materials. Similarly, I have some results as follows:

1. Obviously, in the case of the intercalated metal with low density or being disordered, no periodic lattice of metal positive ions is formed, and the conjugated electron cloud may be localized around the metal ions with some degree of ionic bonding. If the electronegativity of intercalated element and carbon π bond is evenly matched, some covalent feature would appear and the conductivity would turn low. (intercalated atoms could be treated as electron scattered centers, which also leads the conductivity decrease, and the material would have negative temperature correlation coefficient.)

2. CDD (videos in the supplement): taking 2DTh100_CDD_100 as an example, new electron orbitals of Th have been formed along Z-axis after the hybridization of the p, d and f orbitals, and they conjugate with the $p_z$ orbital of the carbon network with some degree of covalent bonding, which originates from the inertness of the s electrons (relativistic effect). The $p_z$ electron cloud is initially distributed right above the carbon atoms; after conjugation the electrons can enter more room in the lattice, so the delocalization and conductivity are enhanced. Note the value in the upper left corner in CDD (like 2DU100_CDD_001), Z(min) is a negative value, which means the charge is lost. Therefore, conjugation also causes the electron clouds of other bonds to shrink. Moreover, the distribution of conjugated electron clouds in three-dimensional space is anisotropy.

3. ELF: taking 2DTh100_ELF_001 as an example, the results show that the ELF value at the distribution position of the electron-conjugated cloud is 0.5 (green), indicating that the electrons in this space exhibit some near-free electron gas behavior. For Pu, the ELF is slightly larger than 0.5, indicating that the bonding at this position has some covalent nature, so the electron movement is localized to some extent. The bonding nature of U-doped GICs is alike to that of Pu-doped GICs.

In summary, the TCD, CDD and ELF demonstrate that the electrons of An metals partly transfer to the π orbital of the carbon network, partly enter the new interlayer bands. The former determines the conductivity, the latter generates some covalent property and bonding anisotropy. The conclusions are consistent with the preceding discussion.

### 3.4 | Absorption spectra

I calculated the absorption spectra of An-doped GICs with 50% density and of Th($C_8H_8$)$_2$, U($C_8H_8$)$_2$ and Pu($C_8H_8$)$_2$. Compared both results, it is found that the (electron orbital) conjugation

(of the actinides to the carbon network) makes the absorption edge migrates from the ultraviolet to the visible region, and the curves change from discrete energy levels to continuous bands (Fig. S1). I tried to identify U-doped GICs with various densities and distributions in accordance with those absorption spectra, however the differences are not obvious (Fig. S2). Therefore, angle-resolved polarized Raman spectroscopy was selected[17].

# 4 | Conclusions

In this work, the research blank of GICs doped with actinide elements is tried to be filled in. Three An metals were selected with representative valence electron arrangement: Thorium (Th) ([Rn]$6d^27s^2$), Uranium (U) ([Rn]$5f^36d^17s^2$) and Plutonium (Pu) ([Rn]$5f^67s^2$). I controlled the distribution and density of the intercalated metals, and consider the effect of changing carbon interlayers distance $d$ on the electronic structure, so as to infer their feasible properties under extreme conditions (e.g. high pressure). According to band schemas, those GICs are all metallic. The projected density of states (PDOS) indicates that the metal atoms first undergo hybridization of its s, p, d, and f orbitals, and then bond with the carbon's $p_z$ orbitals of the nearest-neighbor octagons. The electron orbital spin up of C and Th is symmetrical with spin down, so there is no electron spin polarization. However, the s, p, d, and f orbitals of U and Pu all exhibit electron spin polarization, which leads to the magnetic properties of the material and make the p orbits of C appear spin polarization around Fermi level. Also, it is verified that the room/distance between the actinide metal atoms and the carbon layers strongly affects the electronic structure of the GICs, which originates from the anisotropy of their p, d, and f orbitals.

However, key experimental and theoretical questions remain unanswered at present:

1) Experimentally, the diffusion mechanism of intercalated atoms M into the graphite crystal is unclear and there is no mature, controllable and effective preparation technique. So the obtained GICs' structures were quite random and even unable to be predetermined. There are two means to dope the M atoms: chemical and physical. The former uses halogens, etc. to form bonds at the edge of carbon layer, the latter is by way of external high pressure or electrochemical methods. Either way, diffuse M into the interior of graphite from the edge, so that it inevitably causes uneven density distribution and yet the carbon layer's falling off.

2) Theoretically, the BCS model can successfully describe and predict the superconductivity of bulk GICs doped with alkali or alkaline earth metals; after considering the electron-phonon coupling (EPC), the calculations (e.g. $T_C$) are in the same order of magnitude as the (experimentally) measured data. However, there is no clear and accepted definition for judging whether the electronic structure of a GICs is three-dimensional or two-dimensional. For GICs with two-dimensional Fermi surfaces or with d, f orbital electrons, the relativistic effect cannot be ignored. So how does the electron behavior and the phonon vibration couple in those materials? How does EPC affect the superconductivity? Do new theoretical models need to explore adequately the dependence of $T_C$ on pressure and the distribution, density and chemical composition of the intercalated atom? These long-standing fundamental questions are still desired to be answered.

**ACKNOWLEDGEMENTS**

The author thanks the Physical Research Platform (PRP) in School of Physics, SYSU. Also, thanks for the valuable advice from Professor Zhibing Li and Associate Professor Weiliang Wang.

**DATA AVAILABILITY STATEMENT**

The data that support the findings of this study are available from the corresponding author upon reasonable request.

**NOTES AND REFERENCES**

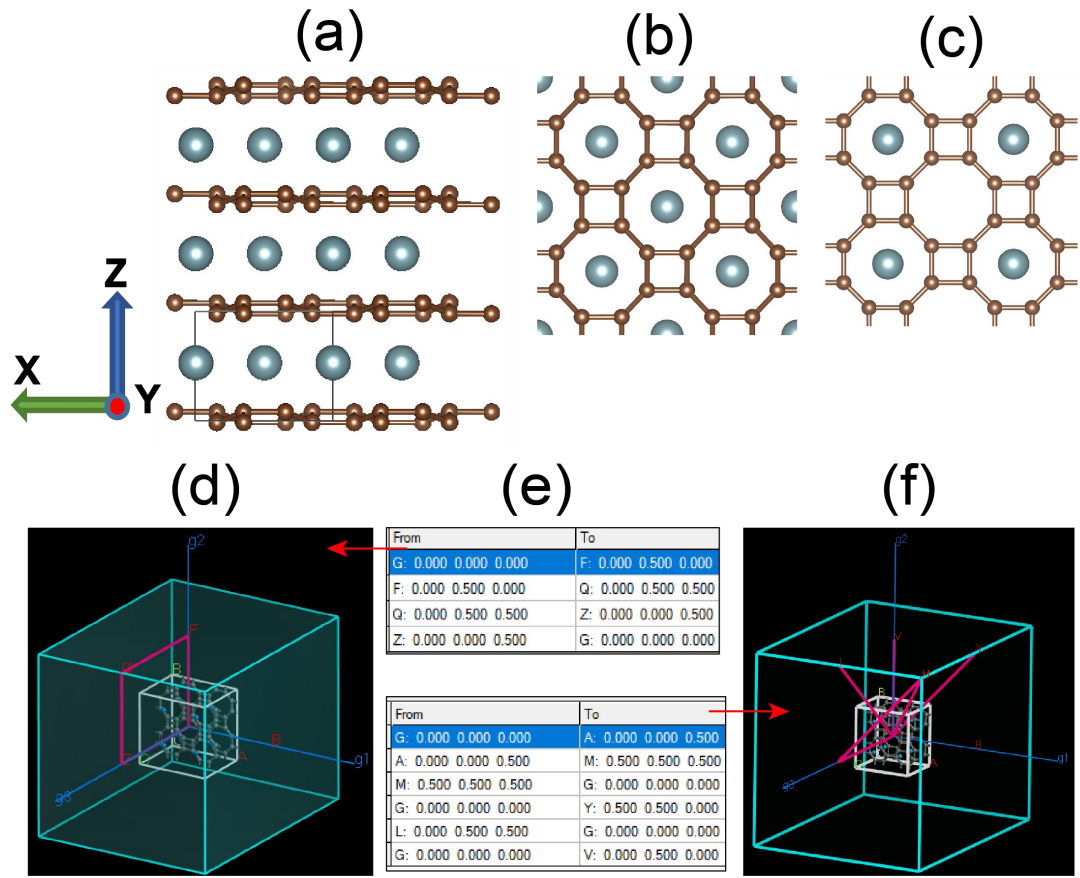

Fig. 1 (a) Stage I GICs viewed along Y-axis; (b) and (c) 2DU100 and 2DU50, respectively (viewed along Z-axis). The blue balls represent metal atoms, and the brown balls represent carbon atoms. (d) is a schematic diagram of the inverse space of GICs (except 2DTh50) and (f) is a schematic diagram of the inverse space of 2DTh50, where white lines represent the lattice in real space and fuchsia lines represent the band paths. (e) The upper table is labels in the band path of (d), and the lower table is labels in the band path of (f).

Table 1 The distance between adjacent carbon layers in Stage I GICs with different metal elements and distribution densities

| metal elements | | distance between adjacent carbon layers ($d$) /Å |
|---|---|---|
| Th | 50% | 4.190 |
|  | 100% | 4.238 |
| U | 50% | 4.123 |
|  | 100% | 3.850 |
| Pu | 50% | 4.053 |
|  | 100% | 4.093 |

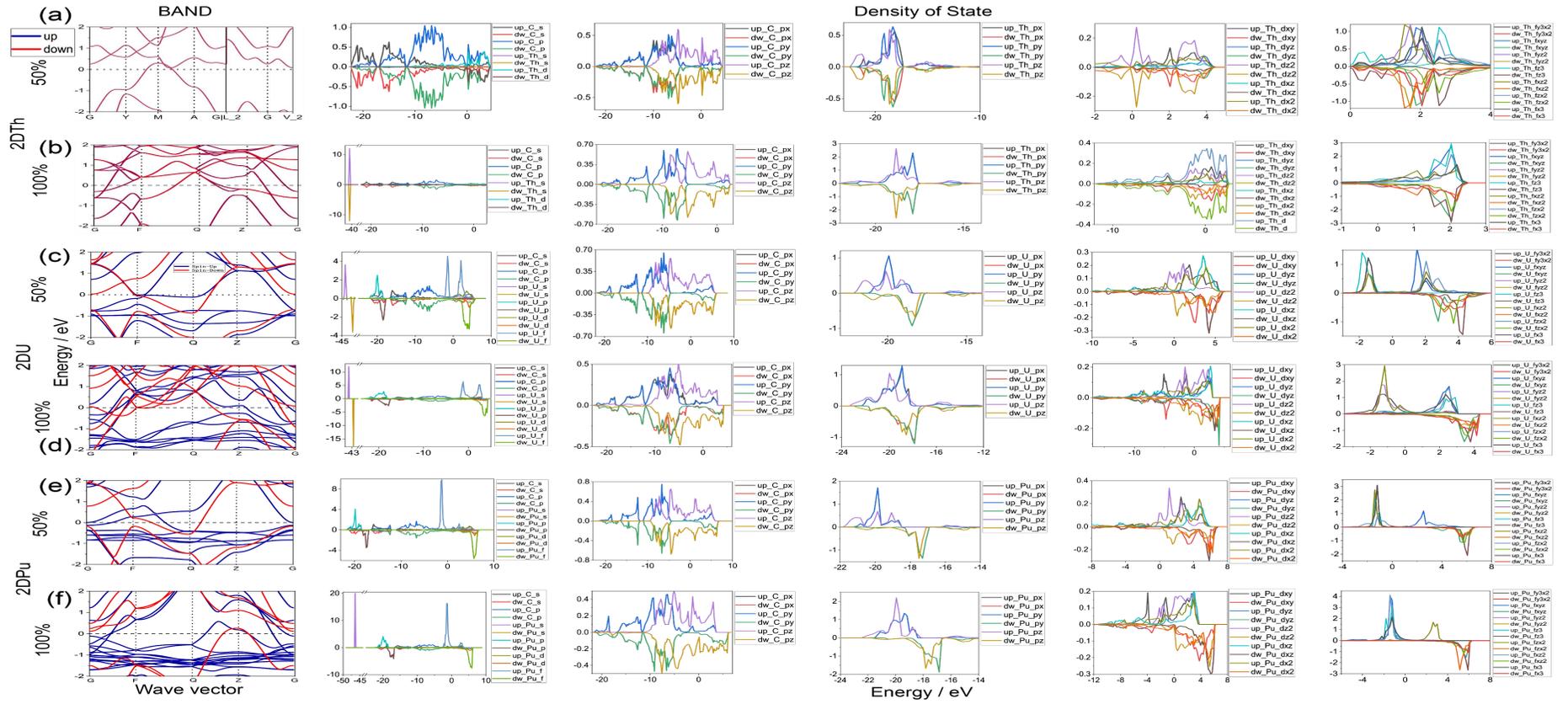

Fig. 2 The electron bands (the leftmost column) and the projected density of states (PDOS) of the GICs doped with Th (a, b), U (c, d) and Pu (e, f) (the second to 6$^{th}$ column). (a, c, e) represent 50% doping density (, (b, d, f) represent 100% doping density. In the band diagram, the blue line represents spin up and the red line represents spin down.

Table 2 d or f orbitals of the An elements that contribute to the DOS near Fermi surface.

|    |      | dxy | dyz | dz2 | dxz | dx2 | fy3x2 | fxyz | fyz2 | fz3 | fxz2 | fzx2 | fx3 |
|----|------|-----|-----|-----|-----|-----|-------|------|------|-----|------|------|-----|
| Th | 50%  |     |     | √   |     |     |       |      |      |     |      |      |     |
| Th | 100% |     |     | √   |     |     |       |      |      |     |      |      |     |
| U  | 50%  |     |     | √   |     |     |       |      |      |     |      |      |     |
| U  | 100% | √   | √   | √   |     | √   | up    |      | up   |     | up   |      |     |
| Pu | 50%  |     |     | up  |     | up  |       |      |      |     | up   |      |     |
| Pu | 100% | √   |     | √   | √   | √   | up    | up   | up   |     | up   |      |     |

* up or dw indicates that the spin up or spin down orbitals contribute, respectively. √ indicates that both spin up and spin down orbitals contribute.

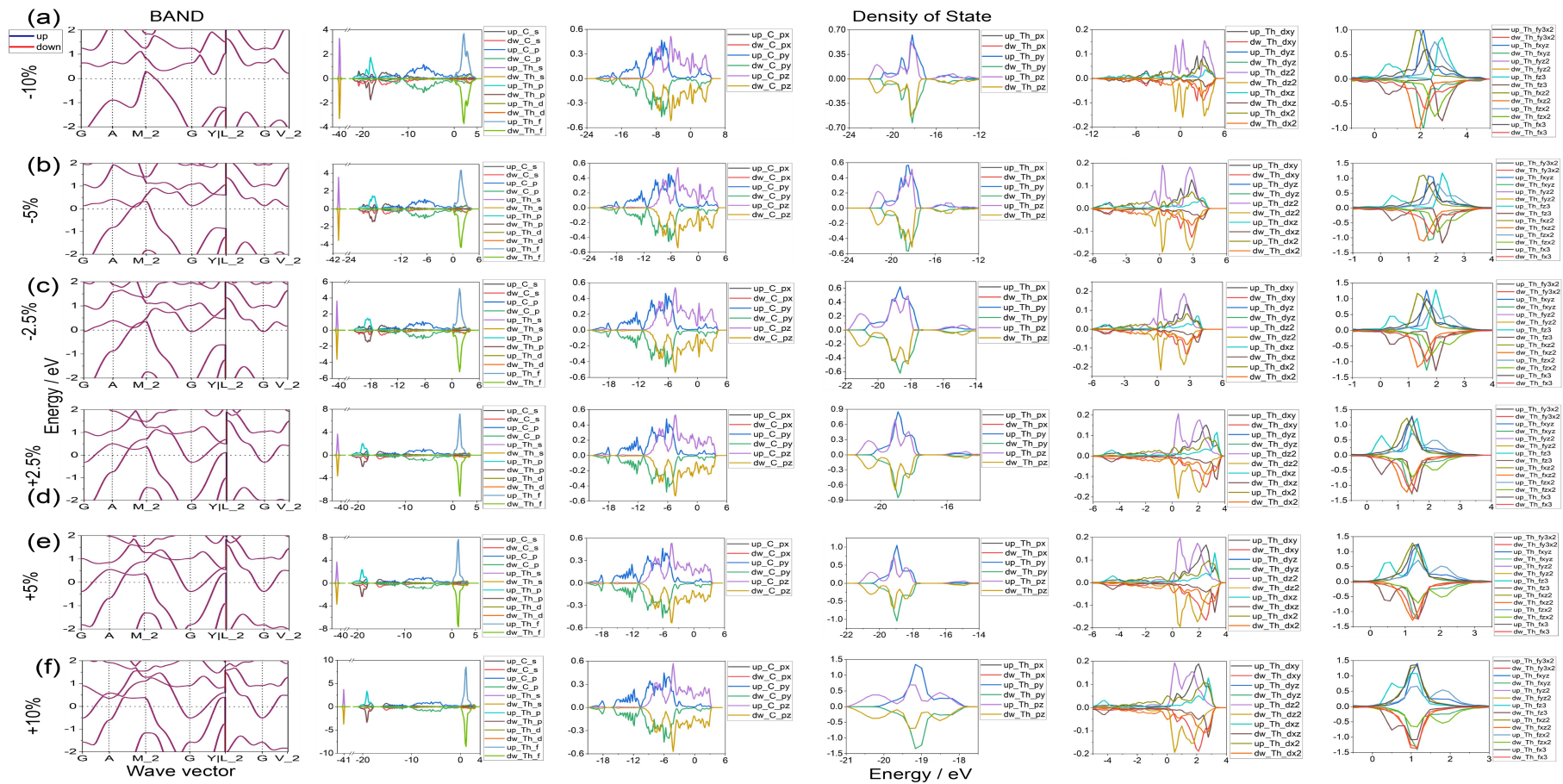

Fig. 3 The shape-change effect on band and PDOS of **2DTh50**. The deformation of the *c*-basal vector is original 90% ((a)-10%), 95% ((b)-5%), 97.5% ((c)-2.5%), 102.5% ((d)+ 2.5%), 105% ((e)+5%), 110% ((f)+10%).

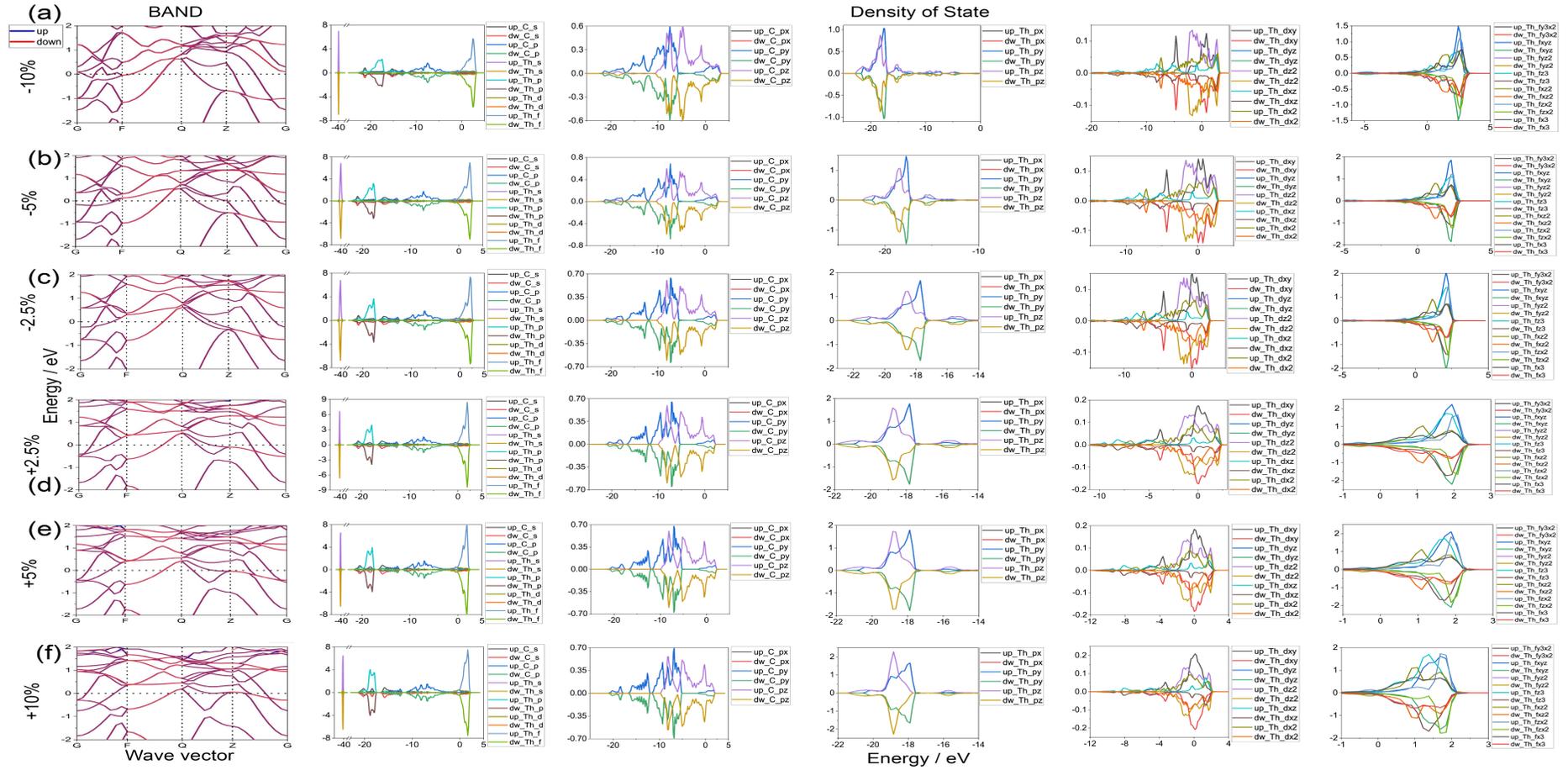

Fig. 4 The shape-change effect on band and PDOS of **2DTh100**. The deformation of the *c*-basal vector is original 90% ((a)-10%), 95% ((b)-5%), 97.5% ((c)-2.5%), 102.5% ((d)+ 2.5%), 105% ((e)+5%), 110% ((f)+10%).

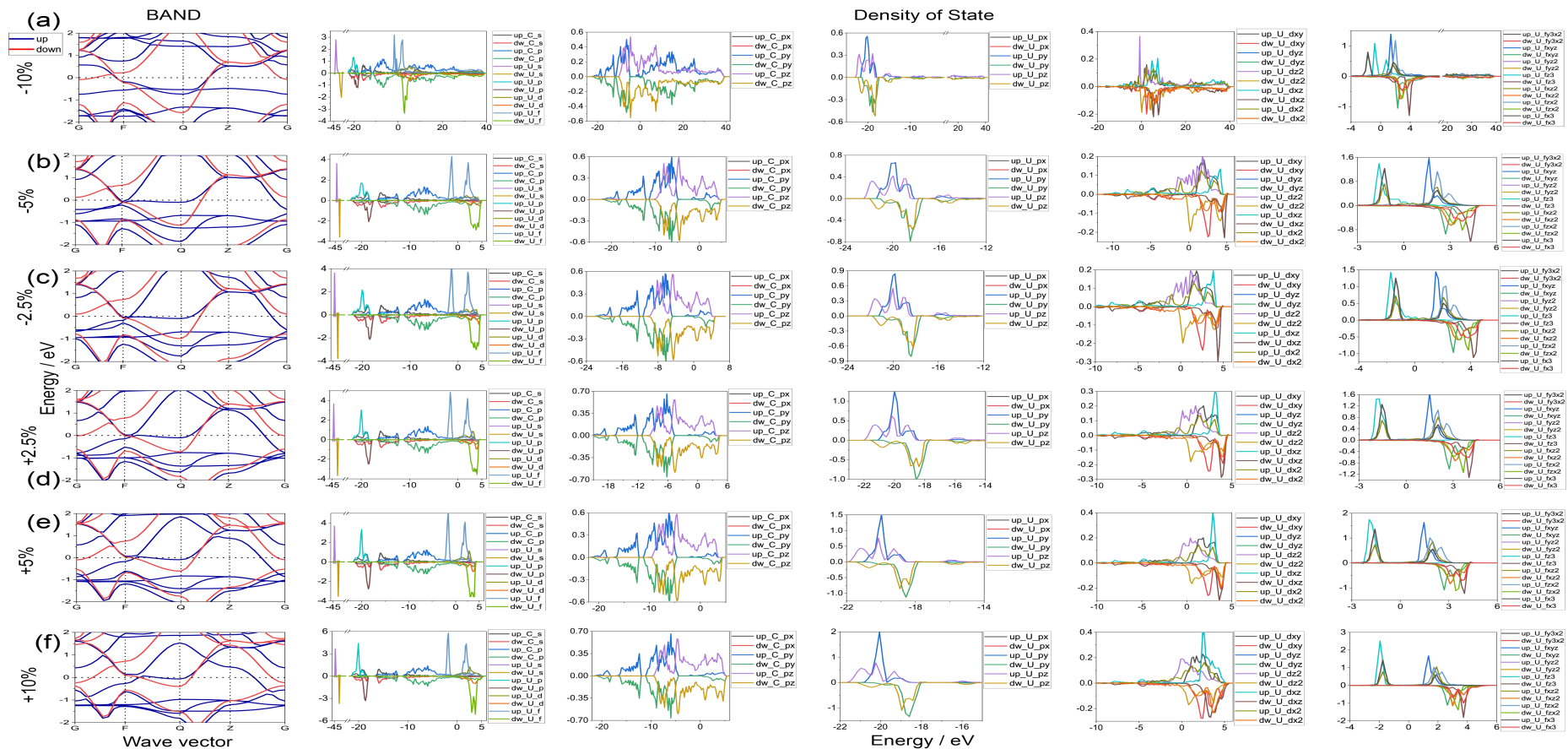

Fig. 5 The shape-change effect on band and PDOS of **2DU50**. The deformation of the *c*-basal vector is original 90% ((a)-10%), 95% ((b)-5%), 97.5% ((c)-2.5%), 102.5% ((d)+ 2.5%), 105% ((e)+5%), 110% ((f)+10%).

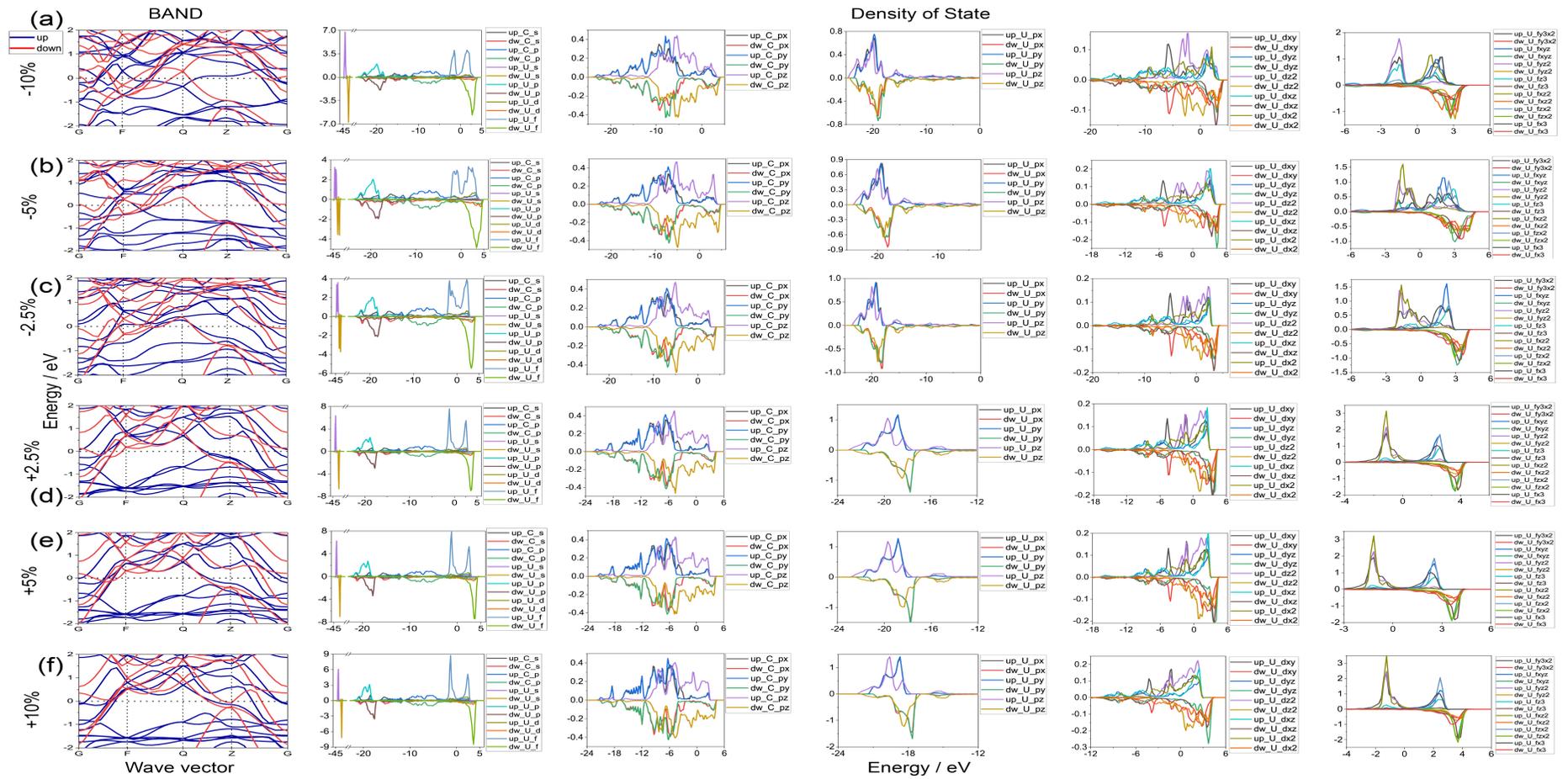

Fig. 6 The shape-change effect on band and PDOS of **2DU100**. The deformation of the *c*-basal vector is original 90% ((a)-10%), 95% ((b)-5%), 97.5% ((c)-2.5%), 102.5% ((d)+ 2.5%), 105% ((e)+5%), 110% ((f)+10%).

Fig. 7 The shape-change effect on band and PDOS of **2DPu50**. The deformation of the *c*-basal vector is original 90% ((a)-10%), 95% ((b)-5%), 97.5% ((c)-2.5%), 102.5% ((d)+ 2.5%), 105% ((e)+5%), 110% ((f)+10%).

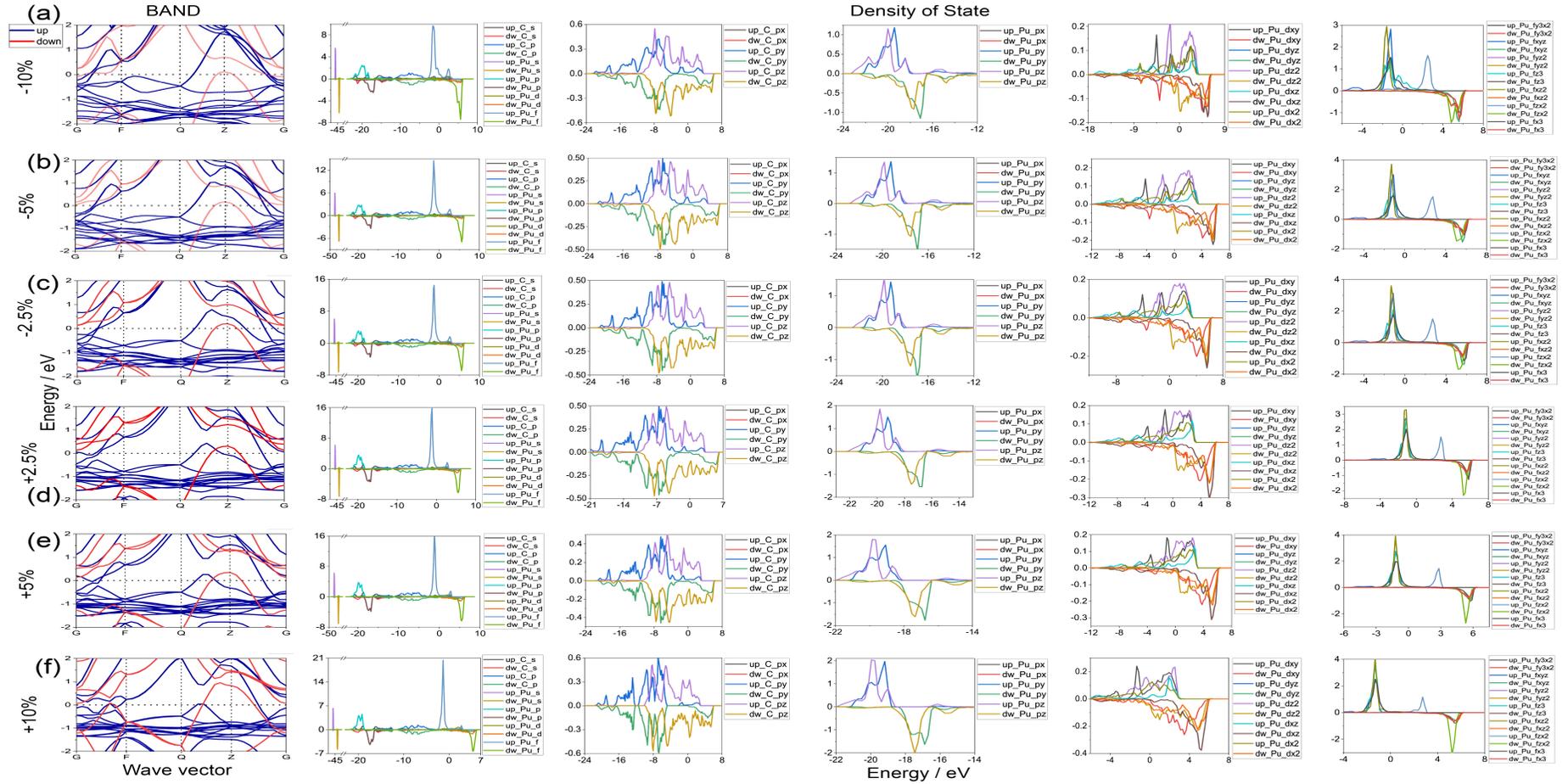

Fig. 8 The shape-change effect on band and PDOS of **2DPu100**. The deformation of the *c*-basal vector is original 90% ((a)-10%), 95% ((b)-5%), 97.5% ((c)-2.5%), 102.5% ((d)+ 2.5%), 105% ((e)+5%), 110% ((f)+10%).

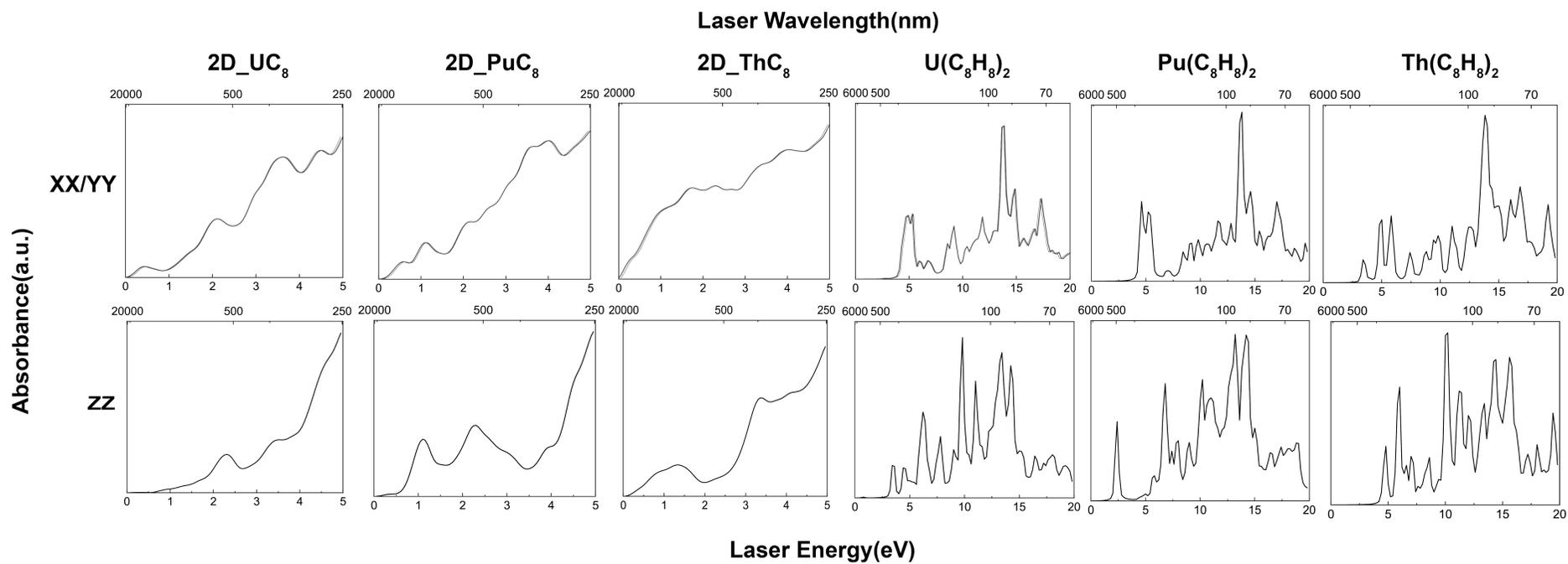

Fig. S1 Absorption spectra of An-doped GICs with 50% density and of $Th(C_8H_8)_2$, $U(C_8H_8)_2$ and $Pu(C_8H_8)_2$

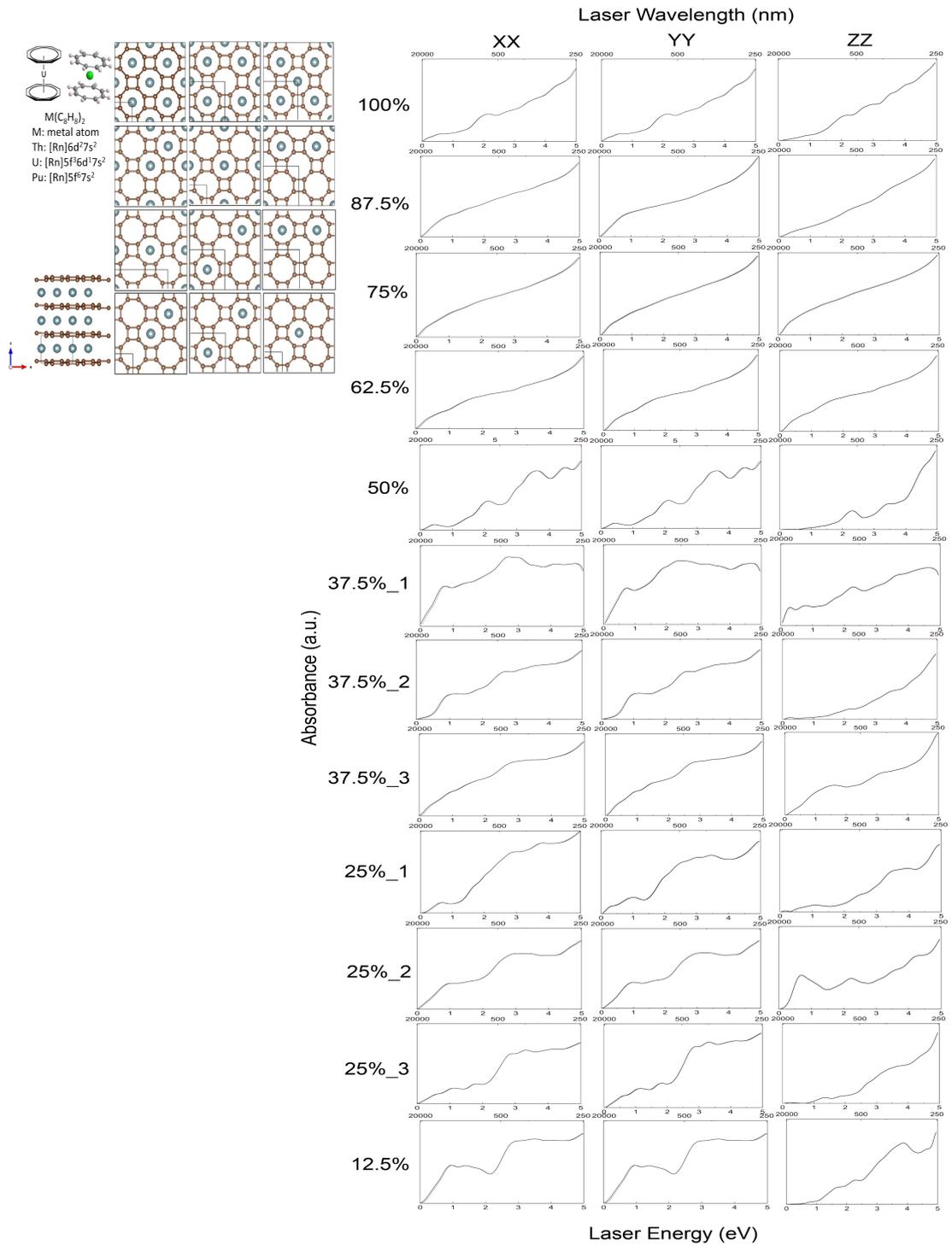

Fig. S2 Structure diagrams and absorption spectra of U-doped GICs with various densities.